\documentclass[twocolumn,showpacs,preprintnumbers,amsmath,amssymb]{revtex4}
\usepackage{epsfig,psfrag}
\usepackage{dcolumn}
\usepackage{bm}
\usepackage{graphicx}
\usepackage{color}
\begin{document}
\title{Full counting statistics for the Kondo dot in the unitary limit}
\author{A. O.~Gogolin$^1$ and A.~Komnik$^2$}
\affiliation{
${}^{1}$ Department of Mathematics, Imperial College London,
180 Queen's Gate, London SW7 2AZ, United Kingdom\\
${}^2$ Service de Physique Th\'{e}orique, CEA Saclay, F--91191
Gif-sur-Yvette, France}
\date{\today}
\begin{abstract}
We calculate the charge transfer probability distribution function
$\chi(\lambda)$ for the Kondo dot in the strong coupling limit
within the framework of the Nozi\`{e}res--Fermi--liquid theory of
the Kondo effect. At zero temperature, the ratio of the moments
$C_n$ of the charge distribution to the backscattering current
$I_{{\rm bs}}$ follows a universal law $C_n/2I_{{\rm
bs}}=(-1)^n(1+2^n)/6$. The functional form of $\chi(\lambda)$ is
consistent with tunnelling of electrons and, possibly, electron
pairs. We then discuss the cross-over behaviour of $\chi(\lambda)$
from weak to strong Coulomb repulsion in the underlying Anderson
impurity model and relate this to the existing results. Finally,
we extend our analysis to the case of finite temperatures.
\end{abstract}
\maketitle
The importance of the noise spectra in charge transport has been
recognised long ago \cite{Schottky}. The higher moments of the
charge distribution, however, are only just becoming accessible
experimentally \cite{Reulet}. The probability $P(Q)$ of charge $Q$
being transmitted through the system during the measuring time
${\cal T}$ and the respective moment generating function
$\chi(\lambda)= \sum_Qe^{i\lambda Q}P(Q)$ have been established by
Levitov and Lesovik \cite{Levitov} for non-interacting electrons.
The effects of electron--electron correlations on the counting
statistics are currently a subject of intensive debate.
In particular, a universal result that the linear
response, zero--temperature statistics is binomial for any
interacting system with one conducting channel has recently
emerged \cite{Gogolin}.

An important test system of such type is the so--called Kondo dot:
a semiconductor quantum dot device in the Kondo regime. Extensive
experimental \cite{Goldbaher-Gordon} and theoretical
\cite{Kaminski,Pustilnik} inquiries into conducting properties of
these systems have recently taken place. In the last few weeks,
two works appeared, Ref.~\cite{Sela} and Ref.~\cite{Golub}, where
the zero--temperature shot noise is calculated for such devices in
the strong--coupling limit.

In this Letter we widen the investigation and calculate the full
charge distribution function (which encodes all the moments) in
the strong--coupling Kondo regime. We also discuss the temperature
corrections and make contact with the previous results for the
Anderson impurity model \cite{Gogolin}.

A good place to start is indeed the Hamiltonian for a single level
quantum dot (which is a version of the Anderson impurity model)
\begin{equation}\label{AndHam}
H=H_0+H_{{\rm T}}+ H_{{\rm C}}\;,
\end{equation}
where
\begin{equation}\label{AndHam0}
H_0=\sum\limits_{p,\sigma,\kappa}(\varepsilon_p-\mu_\kappa)
\psi^{\dagger}_{p,\sigma,\kappa}\psi^{\phantom{\dagger}}_{p,\sigma,\kappa}
+\sum\limits_{\sigma}(\Delta_0+\sigma h)d^\dagger_\sigma
d^{\phantom{\dagger}}_\sigma\;,
\end{equation}
describes the conducting leads, $\psi^{\dagger}_{p,\sigma,\kappa}$
being the creation operator for the electron with momentum $p$,
spin $\sigma=\uparrow,\downarrow$, in the lead $\kappa=R/L$, with
dispersion relation $\varepsilon_p$ such that the Fermi  level
density of states $\nu=\sum_p\delta(\varepsilon_p)$ is finite (we
set $E_F=0$ and $e=\hbar=k_B=1$ throughout), and the fermionic
level $d^\dagger_\sigma$ at the energy $\Delta_0$ in magnetic
field $h$. The leads are DC biased $\mu_L=-\mu_R=V/2\geq 0$.
Further,
\begin{equation}\label{AndHamT}
H_{{\rm T}}=\sum\limits_\sigma\left(\gamma_Ld^\dagger_\sigma
\psi^{\phantom{\dagger}}_{\sigma,L}+\gamma_R\psi^{\dagger}_{\sigma,R}
d^{\phantom{\dagger}}_\sigma+{{\rm H.c.}}\right)
\end{equation}
represents the local electron tunnelling between the leads and the
dot (with, in general, different amplitudes $\gamma_L$ and
$\gamma_R$) while
$H_{{\rm C}} =Un_\uparrow n_\downarrow$
with $n_\sigma=d^\dagger_\sigma d^{\phantom{\dagger}}_\sigma$,
stands for the Coulomb repulsion $U$. In equilibrium, this
probably is the best studied model in the condensed matter theory.
In particular, in the Kondo regime ($\Delta_0<0$, $U$ positive and
large), the Schrieffer--Wolf type transformation
\cite{Schrieffer}, tailored to the lead geometry \cite{Kaminski},
can be applied to result into a Kondo type or s-d exchange model.

For the Kondo model in the strong--coupling limit, when the spin
on the dot is absorbed into the Fermi sea forming a singlet,
Nozi\`{e}res \cite{Nozieres} devised a Landau--Fermi--liquid
description based on a `molecular field' expansion of the phase
shift of the s--wave electrons:
\begin{equation}\label{phase}
\delta_\sigma(\varepsilon)=\delta_0+\alpha\varepsilon+
\phi^{a}(n_\sigma-n_{\bar{\sigma}}) \;,
\end{equation}
where $\delta_0=\pi/2$, $\alpha$ and $\phi^a$ are phenomenological
parameters corresponding to the residual potential scattering and
the residual interactions, respectively. These processes are
generated by polarising the Kondo singlet and so are of the order
$\sim 1/T_K$, where $T_K$ is the Kondo temperature. The specific
heat coefficient is proportional to  $\alpha/(\pi\nu)$ while the
magnetic susceptibility is proportional to the sum
$\alpha/(\pi\nu)+2\phi^a/\pi$. Simple arguments were advanced in
Ref.~\cite{Nozieres} to the effect that, because the Kondo
singularity is tied up to the Fermi level, there exists a relation
$\alpha=2\nu\phi^a$ between the two processes in
Eq.~(\ref{phase}). In particular, this explains why the Wilson
ratio is equal to $2$. Clearly, it is not difficult to write down
a second quantised Hamiltonian describing the processes in
Eq.~(\ref{phase}). Before doing so let us remark that it is well
known that there is no critical $U$ in the Anderson model
(\ref{AndHam}) and the Fermi liquid approach is actually valid for
all $U$ (see \cite{Hewson} for a review).

The strong--coupling Hamiltonian that describes the scattering and
interaction processes encoded in Nozi\`{e}res Eq.~(\ref{phase}) is
of the form $H=H_0+H_{{\rm sc}}+ H_{{\rm int}}$. The free
Hamiltonian here is
\begin{equation}\label{Hfree}
H_0=\sum\limits_{p,\sigma}\varepsilon_p(c^\dagger_{p,\sigma}c^{\phantom{\dagger}}_{p,\sigma}
+a^\dagger_{p,\sigma}a^{\phantom{\dagger}}_{p,\sigma})+VQ\;,
\end{equation}
where $c^{\dagger}$ is the creation operator for the s--wave
electrons, $a^{\dagger}$ is the creation operator of the p--wave
electrons, included in order to account for the transport
\cite{Pustilnik}, and the operator
\[
Q=\frac{1}{2}\sum\limits_{p,\sigma}(c^\dagger_{p,\sigma}
a^{\phantom{\dagger}}_{p,\sigma}
+a^\dagger_{p,\sigma}c^{\phantom{\dagger}}_{p,\sigma})\;
\]
stands for the (minus) charge transferred across the junction.
The scattering term is
\begin{equation}\label{Hsc}
H_{{\rm sc}}=\frac{\alpha}{2\pi\nu T_K}\sum\limits_{p,p',\sigma}
(\varepsilon_p+\varepsilon_{p'})c^\dagger_{p,\sigma}c^{\phantom{\dagger}}_{p',\sigma}\;,
\end{equation}
while the interaction term reads
\begin{equation}\label{Hint}
H_{{\rm int}}=\frac{\phi}{\pi\nu^2 T_K}
c^\dagger_{\uparrow}c^{\phantom{\dagger}}_{\uparrow}
c^\dagger_{\downarrow}c^{\phantom{\dagger}}_{\downarrow}\;,
\end{equation}
where $c^{\phantom{\dagger}}_{\sigma}=\sum_pc^{\phantom{\dagger}}_{p,\sigma}$
and we have changed to the dimensionless amplitudes $\alpha$ and $\phi$,
so that in the actual Kondo model $\alpha=\phi=1$ (in the intermediate
calculations it is convenient to treat $\alpha$ and $\phi$ as free parameters
though). By the nature of the strong--coupling fixed point,
the operators $\alpha$ and $\phi$ are irrelevant in the renormalisation
group sense and therefore the perturbative expansion in $\alpha$ and $\phi$
is expected to converge.

The charge {\it measuring field} $\lambda(t)$
[$\lambda(t)=\lambda\theta (t)\theta({\cal T}-t)$ on the forward
path and $\lambda(t)=-\lambda\theta (t)\theta({\cal T}-t)$ on the
backward pass of the Keldysh contour $C$], couples, in the
Lagrangian formulation, to the current \cite{Levitov2} via a term
in the action $\int dt \lambda(t) \dot{Q}(t)=-\int dt
\dot{\lambda}(t)Q(t)$, which can be gauged away by the canonical
transformation
\begin{equation}\label{rotation}
\begin{array}{llll}
c\to c_\lambda
& = & \cos(\lambda/4) c- i \sin(\lambda/4) a\;,\\
a\to a_\lambda
& = & -i\sin(\lambda/4) c+ \cos(\lambda/4) a\;.
\end{array}
\end{equation}
We therefore reach an important conclusion that the charge
measuring field enters this problem as a rotation of the
strong--coupling basis of the s-- and the p--states. While $H_0$
is invariant under this substitution, it should be performed in
both the scattering and the interaction Hamiltonians, $H_{{\rm
sc}}[c]+H_{{\rm int}}[c] \to H_\lambda=H_{{\rm
sc}}[c_\lambda]+H_{{\rm int}}[c_\lambda]$, when calculating the
statistics. It is easily checked that at the first order in
$\lambda$: $H_\lambda=H_{{\rm sc}}+ H_{{\rm int}}
+(\lambda/4)\hat{I}_{{\rm bs}}+O(\lambda^2)$, where $\hat{I}_{{\rm
bs}}$ is the backscattering current operator
\begin{eqnarray}
\hat{I}_{{\rm bs}}=& -&i\frac{\alpha}{4\pi \nu T_K}\sum\limits_{p,p',\sigma}
(\varepsilon_p+\varepsilon_{p'})(c^\dagger_{p,\sigma}a^{\phantom{\dagger}}_{p',\sigma}
-a^\dagger_{p,\sigma}c^{\phantom{\dagger}}_{p',\sigma})\nonumber\\
&-&i\frac{\phi}{2\pi\nu^2 T_K}\sum\limits_{\sigma}
(c^\dagger_{\sigma}a^{\phantom{\dagger}}_{\sigma}-
a^\dagger_{\sigma}c^{\phantom{\dagger}}_{\sigma})
c^\dagger_{\bar{\sigma}}c^{\phantom{\dagger}}_{\bar{\sigma}}\;,
\label{current}
\end{eqnarray}
alternatively available from the commutator
$\hat{I}_{{\rm bs}}=-\dot{Q}=i[Q,H]$.

The charge probability distribution function is therefore defined
by
\begin{equation}\label{chidef}
\chi(\lambda)=e^{i N \lambda} \left \langle T_C \exp\left[-i\int_C
H_\lambda(t) dt\right] \right\rangle \; ,
\end{equation}
where $T_C$ is the time ordering
operator on the Keldysh contour, and $N=(V{\cal T})/\pi = (2e^2/h) V {\cal
T}$. Applying the standard linked cluster expansion (still valid
on the Keldysh contour, of course)\cite{AGD}, we see that the
leading correction to the distribution function is given by a
connected average
\begin{equation}\label{formalcorr}
\ln\chi(\lambda)=i N \lambda-\frac{1}{2}\int_C dt_1dt_2\langle
T_C\{H_\lambda(t_1)H_\lambda(t_2)\}\rangle_{{\rm c}}+...
\end{equation}
The neglected terms $\alpha^4$, $\alpha^2\phi^2$, $\phi^4$, etc.,
are of the higher order in voltage (temperature) than the main
correction because of the irrelevant nature of the perturbation.
In order to make progress with  Eq.~(\ref{formalcorr}), one only
needs the Green's function of the $\lambda$--rotated
$c$--operator, which is easily seen to be the following matrix in
Keldysh space
\begin{widetext}
\begin{eqnarray}
\hat{g}_\lambda(p,\omega)=i\pi\delta(\varepsilon_p-\omega)
\left\{\right.[f(\omega-V/2)+f(\omega-V/2)-1]\hat{\tau}_0
+[e^{-i\lambda/2}f(\omega-V/2)+e^{i\lambda/2}f(\omega+V/2)]
\hat{\tau}_+\label{gfbare} \\
-[(1-f(\omega-V/2))e^{i\lambda/2}+(1-f(\omega+V/2))e^{-i\lambda/2}]
\hat{\tau}_- \left. \right\}\;,\nonumber
\end{eqnarray}
\end{widetext}
where $\hat{\tau}_i$ is the standard choice of Pauli matrices and
$f(\omega)$ is the Fermi distribution function. We did not write
the principal part which does not contribute to local quantities
in the flat band model \cite{Lifshits}.

\begin{figure}
\epsfig{file=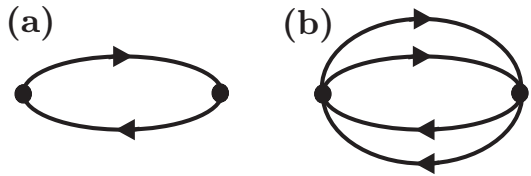,height=2.25cm}
\caption[]{\label{diagrams} Schematic representation of the
scattering (a), and interaction contributions (b).}
\end{figure}

The correction to the distribution function which is due to the
scattering term (\ref{Hsc}) is shown in Fig.~\ref{diagrams}(a).
The corresponding analytic expression is (a factor of 2 comes from
summing over the spin index)
\begin{widetext}
\begin{eqnarray}
\delta_\alpha \ln \chi (\lambda) &=&
 -\frac{\alpha^2}{4\pi^2\nu^2
T_K^2}\sum\limits_{p_1,p_2}(\varepsilon_{p_1}+
\varepsilon_{p_2})^2 \int_C dt_1
dt_2 g_{p_2}(t_2,t_1)
g_{p_1}(t_1,t_2)
\nonumber\\
&=& \frac{  \alpha^2{\cal T}}{\pi T_K^2}\int
d\omega\omega^2[(e^{-i\lambda}-1)n_L(1-n_R)+
(e^{i\lambda}-1)n_R(1-n_L)]\;,\label{alphaT}
\end{eqnarray}
\end{widetext}
which, at zero temperature, contributes to Eq.~(\ref{formalcorr})
a term
\begin{equation}\label{alphacorr}
\delta_\alpha \ln \chi (\lambda) = \frac{ \alpha^2V^3{\cal
T}}{12\pi T_K^2}(e^{-i\lambda}-1)\;.
\end{equation}

Regarding the correction to the charge distribution coming from
the interaction term (\ref{Hint}), any diagrams with a single
insertion of the Green's function vanish (therefore there is also
no $\alpha\phi$ cross term) and the only remaining connected graph
is shown in Fig.~\ref{diagrams}(b). This is best calculated in
real time. At zero temperature we have (no summation over the spin
index):
\begin{eqnarray}
&& \delta_\phi \ln \chi (\lambda) = -\frac{\phi^2}{2\pi^2\nu^4
T_K^2}\int_C dt_1 dt_2 g(t_1,t_2)^2 g(t_2,t_1)^2
\label{phiT} \nonumber \\
&=&~\frac{\phi^2}{\pi^2 T_K^2} \int\limits_{-\infty}^{\infty}dt
\frac{\cos^4[\lambda/2+(Vt)/2)]}{(t+i\alpha)^4}
\\
&=&~\frac{\phi^2 V^3{\cal T}}{12\pi T_K^2}(e^{-i\lambda}-1)+
\frac{\phi^2 V^3{\cal T}}{6\pi
T_K^2}(e^{-2i\lambda}-1)\;.\nonumber
\end{eqnarray}

Combining the results we find that the zero--temperature charge
distribution function is of the form:
\begin{eqnarray}\label{fullcorr}
\ln\chi(\lambda)&=& i N \lambda+\frac{V^3{\cal T}}{12\pi T_K^2}
\left[(\alpha^2+\phi^2)(e^{-i\lambda}-1)
  \nonumber  \right. \\  &+& \left. 2\phi^2(e^{-2i\lambda}-1)\right]+O(V^5)\;.
\end{eqnarray}
In particular, the results of \cite{Golub} for the average
backscattering current and noise power are correctly reproduced
(there the notation is $a=\alpha/(2\pi), b=\phi/\pi$).
Therefore the Fano factor is equal to
\begin{equation}\label{Fano}
\frac{C_2}{2I_{{\rm bs}}}=\frac{\alpha^2+9\phi^2}{2(\alpha^2+5\phi^2)}\;,
\end{equation}
which is $1/2$ for non-interacting electrons ($\phi=0$) and
$5/6$ at the Kondo fixed point ($\alpha=\phi$).

But we now know much more -- all the moments can be computed from
Eq.~(\ref{fullcorr}) via
$C_n=(-i)^n (d^n/d\lambda^n)\ln\chi(\lambda)|_{\lambda=0}$,
leading to
\begin{equation}\label{moments}
\frac{C_n}{2I_{{\rm bs}}}=(-1)^n\frac{\alpha^2+(1+2^{n+1})\phi^2}{2
(\alpha^2+5\phi^2)}\;,
\end{equation}
$n=2,3,...$. Hence at $\alpha=\phi=1$ we have a hierarchy
of \emph{universal} (Fano factor inspired) ratios
$C_n/2I_{{\rm bs}}=(-1)^n(1+2^n)/6$.

In some systems, like in the case of the fractional quantum Hall
effect, there are physical grounds on which to expect fractional
charge quasiparticles to appear in transport \cite{depicciotto}.
In the present problem, which after all answers the Fermi liquid
description, there are no such circumstances. In fact, the
functional form of the distribution function allows a very simple
interpretation. The term proportional to
$V^3(\alpha^2+\phi^2)(e^{-i\lambda}-1)$ must be interpreted as
tunnelling of conventional electrons with charge $e$ (this is
indeed obvious from the fact that at $\phi=0$ there is no
interaction and the problem is trivial). It is tempting to
interpret the term proportional to $V^3\phi^2(e^{-2i\lambda}-1)$,
which appears as a result of electron correlations, as a coherent
tunnelling of electron pairs with charge $2e$. [Certain caution is
required here, as one can show that, at higher orders, a term of
the form $V^5\alpha^4(e^{-2i\lambda}-1)$ exists, which is an
artefact of the expansion around the perfect transmission.]

Let us now return to the original model Eq.~(\ref{AndHam}). There
is an extensive vintage literature on the Fermi liquid properties
of the equilibrium Anderson model \cite{Yamada1}. The perturbative
expansion in the Coulomb repulsion $U$ can often be re-summed to
all orders, using Ward identities and similar techniques, so that
the observable quantities (specific heat, conductance,
magnetisation) are expressed in terms of the even at odd
susceptibilities $\chi_e$ and $\chi_o$ (correlations of
$n_\uparrow$ with $n_\uparrow$ and $n_\downarrow$, respectively).
These are universal functions of $U$ and are known exactly from
the Bethe ansatz \cite{OK,TW}. Specifically, in the weak coupling
$\chi_{{\rm e}}=1+(3-\pi^2/4) U^2/(\pi^2\Gamma^2)+...$
and $\chi_{{\rm o}}=-U/(\pi\Gamma)$,
where $\Gamma=2(\pi\nu\gamma)^2$ (we assume, for simplicity, a
symmetric junction, $\gamma_L=\gamma_R=\gamma$ and particle--hole
symmetry $\Delta_0=-U$). On the other hand, in the strong coupling
we have:
$\chi_e= (\Gamma \alpha)/(\pi T_K)$ and $\chi_o=(\Gamma \phi)/(\pi
T_K)$,
as here $\alpha=\phi=1$ and $T_K$ is the Kondo temperature up to a
prefactor \cite{TW,OK,Hewson}. The programme of extending a Fermi
liquid approach to non-equilibrium properties of the Anderson
model has not been comprehensively carried out yet.

There is a Fermi-liquid proof, due to Oguri \cite{Oguri}, that the
leading non-equilibrium correction to the zero--temperature
current is of the form
\begin{equation}\label{current-Oguri}
I_{{\rm bs}}=\frac{V^3}{12\pi^2\Gamma^2}(\chi_e^2+5\chi_o^2)\;,
\end{equation}
which is valid for all $U$ and interpolates between the
weak--coupling and the strong--coupling regimes of the Anderson
model. We see that the above result for $I_{{\rm bs}}$ is simply the
strong--coupling limit of Oguri's formula.

As to the noise and higher moments no analogous Fermi--liquid
results exist, to the best of our knowledge. However, in our
recent paper \cite{Gogolin}, we did put forward the formula for
the charge distribution function:
\begin{eqnarray}\label{guess}
\ln\chi(\lambda)&=& N\left\{i\lambda+\frac{V^2}{3\Gamma^2} \left[
\frac{\chi_e^2+\chi_o^2}{4}(e^{-i\lambda}-1) \right. \right. \\
\nonumber
 &+& \left. \left. \frac{\chi_o}{2} (e^{-2i\lambda}-1) \right]\right\}\;,
\end{eqnarray}
which was a guess based on the properties of the perturbative
expansion in $U$. In \cite{Gogolin} we have shown that
Eq.~(\ref{guess}) is correct at the order $U^2$ [it is also
consistent with Eq.~(\ref{current-Oguri}) for all $U$].
The calculations in this Letter prove that Eq.~(\ref{guess})
is also correct in the strong--coupling limit.

Before we close, we wish to discuss the effects a finite
temperature $T$. The energy integral in Eq.(\ref{alphaT}) is
elementary. The time integration in Eq.~(\ref{phiT}) can also be
done with standard methods (the real time Green's function at
$T>0$ is related to that at $T=0$ by a well known conformal
mapping $1/t\to\pi T/\sinh(\pi Tt)$, see \cite{book}). The result
for the generating function is:
\begin{widetext}
\begin{eqnarray}
\ln\chi(\lambda)&=& i N \lambda +  \frac{ V{\cal T}}{24 \pi T_K^2}
\left\{ \frac{\alpha^2+\phi^2}{\sinh(V/2T)} [V^2+4(\pi
T)^2][e^{V/2T}(e^{-i\lambda}-1)
+e^{-V/2T}(e^{i\lambda}-1)] \right. \nonumber \\
&+& \left. \frac{2 \phi^2}{\sinh(V/T)}[V^2+(\pi
T)^2][e^{V/T}(e^{-2i\lambda}-1)+ e^{-V/T}(e^{-2i\lambda}-1)]
\right\} \; . \label{result}
\end{eqnarray}
Using this formula we recover the finite-$T$ version of
Eq.~(\ref{current-Oguri}), see \cite{Oguri}. The higher moments
generated from Eq.~(\ref{result}), starting with $n=2$, are all
new results (to our knowledge). The most important of these is the
expression for the noise, which we here report:
\begin{eqnarray}
C_2 = \frac{V}{\pi T_K^2}\left\{\coth\left(\frac{V}{2T}\right)
\left[\frac{\alpha^2+\phi^2}{12}V^2+
\frac{\alpha^2+\phi^2}{3}\right. (\pi T)^2\right]
+ \left.\coth\left(\frac{V}{T}\right) \frac{2
\phi^2}{3}\left[V^2+(\pi T)^2\right]\right\}\;.\label{noiseT}
\end{eqnarray}
\end{widetext}

To summarise, we have calculated the full counting statistics near
the unitary strong--coupling limit of the Kondo dot. We wish to
stress that Eq.~(\ref{guess}), even though correct for all the
moments in both weak--coupling and strong--coupling limits, has
not been formally proven yet. The theoretical challenge is to
develop Fermi liquid theory for the counting statistics.

The authors participate in the European network DIENOW. AK is
Feodor Lynen Fellow of the Alexander von Humboldt foundation.

\end{document}